%
%

\baselineskip 14pt plus 2pt

\font\llbf=cmbx10 scaled\magstep2
\font\lbf=cmbx10 scaled\magstep1

\def\ni{\noindent}

\def\ua{\underline a \,}

\def\bi{\bf i \,}
\def\bj{\bf j \,}
\def\bk{\bf k \,}
\def\bl{\bf l \,}

\def\uA{\underline A \,}      
\def\uB{\underline B \,}      
\def\uC{\underline C \,}


\ni
{\llbf On the role of conformal three-geometries in the dynamics of 
General Relativity}\par
\bigskip
\bigskip

\ni
{\bf L\'aszl\'o B. Szabados}\par
\ni
Research Institute for Particle and Nuclear Physics \par
\ni
H-1525 Budapest 114, P.O.Box 49, Hungary\par
\ni
e-mail: lbszab@rmki.kfki.hu\par
\bigskip
\bigskip

\ni
It is shown that the Chern--Simons functional, built in the spinor 
representation from the initial data on spacelike hypersurfaces, is 
invariant with respect to infinitesimal conformal rescalings if and 
only if the vacuum Einstein equations are satisfied. 
As a consequence, we show that in the phase space the Hamiltonian 
constraint of vacuum general relativity is the Poisson bracket of 
the imaginary part of this Chern--Simons functional and Misner's 
time (essentially the 3-volume). Hence the vacuum Hamiltonian 
constraint is the condition on the canonical variables that the 
imaginary part of the Chern--Simons functional be constant along the 
volume flow. The vacuum momentum constraint can also be reformulated 
in a similar way as a (more complicated) condition on the change of 
the imaginary part of the Chern--Simons functional along the flow of 
York's time. 
\bigskip
\bigskip

\ni
{\lbf 1. Introduction}\par
\bigskip
\ni
In the initial value formulation of general relativity the evolution of 
states is described with respect to a {\it topological}, or coordinate 
time, and we can speak about the {\it metrical}, i.e. the physical time, 
e.g. the proper length of the history of a massive particle or of an 
observer, only {\it after} solving the evolution equations (see e.g. 
[1-3]). Although this does not yield any problem in classical physics, 
it yields serious conceptional difficulties in the quantum theory both 
of the matter fields on a curved spacetime and of the gravity itself 
(see e.g. [4-6]). The resolution of this difficulty could be the 
isolation of certain (matter and/or gravitational) degrees of freedom 
as the `natural' time variable (`internal clock'), and the evolution 
of the remaining degrees of freedom would be described with respect 
to this `internal time' variable [6-9]. However, this `internal time' 
would be defined on the {\it phase space} of the physical system 
rather than in the spacetime (`external time'), and hence it is not a 
priori obvious that these two concepts of time should coincide even 
if we have a well defined external time. To illustrate the potential 
difficulties, let us consider the planar rotor, the simplest possible 
model of classical clocks: The position of the pointer is given by an 
angle coordinate $\varphi\in[0,2\pi)$, and we say that the clock is 
running if $\dot\varphi$, the derivative of the `internal time' 
$\varphi$ with respect to the `external time' $t$, is strictly 
positive (or negative). Then the actual `internal time' shown by the 
clock is not only the value of the angle variable $\varphi$, but 
$\varphi$ {\it plus $2\pi$-times the number of full periods the clock 
has taken}. Thus $\varphi$ could be a global `internal time' only if 
$\dot\varphi$, as a function of $\varphi$, tends to zero as $\varphi
\rightarrow2\pi$, i.e. if the clock `slows down to zero' 
asymptotically before taking one complete period. Thus if $\dot\varphi
\geq c>0$ for some constant $c$, then the `internal time' $\varphi$ 
cannot be globally well defined. In fact, by Poincare's recurrence 
theorem [10] this seems to be a general property of any {\it localized, 
quasi-stationary} clock modeled by a classical Hamiltonian mechanical 
system (even with noncompact configuration space): If the dynamics is 
forced to take place in an open subset $W$ with compact closure of the 
phase space (e.g. by a potential increasing monotonically at infinity 
to bound the system's positions and momenta), then for any point $p\in 
W$ and its arbitrarily small open neighbourhood $U$ there is an 
`external time' parameter value $t_{(p,U)}$ such that the system's 
dynamical trajectory through $p$ at $t=0$ returns to $U$ before $t
_{(p,U)}$, and hence there is no continuous function on $W$ which 
would be monotonically increasing along the dynamical trajectory. (If 
the system is not localized, then, of course, one can find such a 
function, e.g. one of the Cartesian coordinates of a free particle.) 
Thus one should think of the `external time', i.e. the natural parameter 
along the dynamical trajectories, only as an abstraction of the 
`internal times' of specific localized clocks, like the manifold, 
which is defined in terms of local Euclidean neighbourhoods, and one 
cannot expect the `internal times' to be global. What one can expect is 
that the {\it derivative} of one with respect to the other be bounded 
from below by a positive constant. 

Returning to general relativity, technically the choice for such 
an `internal clock' would be the fixing of the lapse function of the 
foliation in an intrinsic, geometric way. In fact, in the expanding or 
contracting phase of the Bianchi I. and IX. cosmological models Misner 
[11] found the volume of the hypersurface of the spatial homogeneity to 
be such a natural time coordinate. 
In the basic paper ``Role of conformal three-geometries in the dynamics 
of general relativity'' York showed how the unconstrained (physical) 
degrees of freedom of vacuum general relativity can be characterized by 
the conformal 3-geometry of the spacelike hypersurfaces, and one of the 
canonical momenta, the trace $T:={2\over3\kappa}\chi$ of the extrinsic 
curvature of the spacelike hypersurfaces, could be identified as a 
natural time variable [12]. (Here $\kappa$ is Einstein's gravitational 
constant.) In fact, by the Raychaudhury equation $T$ is monotonic in 
time provided the strong energy condition holds (e.g. in vacuum) and 
the acceleration of the hypersurface can be neglected. Thus it may be 
monotonic even when Misner's time has a turning point. Unfortunately 
$T$ is not monotonic in general either. Recently Smolin and Soo argued 
that since the proper arena of the dynamics is the phase space rather 
than the spacetime and in a canonical quantum theory the carrying space 
of the wave functions is the configuration space, we should find a 
natural time variable in the configuration space and not in the 
spacetime. For such a natural time variable in the configuration space 
they suggested the imaginary part of the Chern--Simons functional built 
from the complex Ashtekar connection [13]. Although the quantum dynamics 
must be formulated in the configuration space (or on the phase space 
endowed with an appropriate polarization), we think that in the {\it 
classical theory} time in the phase space or configuration space (i.e. 
the `internal time') and time in the spacetime (`external time') should 
be monotonic with respect to each other, even if the former is not a 
globally defined observable on the phase- or configuration space. 
Unfortunately, their time variable is also of limited validity [14]. 

These negative results rise the question as whether this failure of 
finding the `internal time' is an indication of the non-existence of 
such a time in the field theoretic framework either, at least in the 
generic case. It is an open question whether the dynamics of the vacuum 
Einstein equations are analogous to that of the free particle, or there 
exists an appropriate version of Poincare's recurrence theorem for 
infinite dimensional Hamiltonian systems with physically reasonable 
conditions that would rule out the existence of `intrinsic times'. 
On the basis of a Machian analysis Barbour goes further [15] saying 
that the idea of the `intrinsic time' and clocks is wrong, and ``any 
satisfactory operational definition of time must involve all the 
degrees of freedom of the universe on an equal footing'', and it is 
legitimate to use clocks {\it only} in the description of the dynamics 
of the {\it subsystems} of the whole universe. 

The Chern--Simons functional is known to play an interesting role 
in the characterization of the conformal structures on 3-manifolds. 
In fact, the Chern--Simons functional built from the Levi-Civita 
connection of a Riemannian 3-manifold $(\Sigma,q_{\mu\nu})$ is 
invariant with respect to conformal rescalings of the 3-metric $q_{\mu
\nu}$ [16]. Furthermore, in the Lichnerowitz--Choquet-Bruhat--York 
(conformal) approach of solving the constraint equations for the 
constant mean curvature data set of general relativity it is only 
the {\it conformal class} of the 3-metric, and the (unphysical) 
transverse--traceless extrinsic curvature and the trace of the physical 
extrinsic curvature that should be prescribed [17]. But the curvature 
of the conform 3-geometries is the Cotton--York tensor, which is just 
the variational derivative of the Chern--Simons functional above with 
respect to the metric [16]. Thus some role of the Chern--Simons 
functional in the dynamics of the 3+1 dimensional general relativity 
may be expected. In some of our previous papers we gave two 
generalizations of this Chern--Simons conformal invariant for triples 
$(\Sigma,q_{\mu\nu},\chi_{\mu\nu})$, where $\chi_{\mu\nu}$ is any 
symmetric tensor field on $\Sigma$ [18]. The first was based on a real 
Lorentzian vector bundle over $\Sigma$ and a covariant derivative 
thereon, built from $q_{\mu\nu}$ and $\chi_{\mu\nu}$. This is invariant 
with respect to transformations of $q_{\mu\nu}$ and $\chi_{\mu\nu}$ 
corresponding to {\it spacetime} conformal rescalings. The second 
generalization was based on the {\it complex} vector bundle of 
anti-self-dual 2-forms, but that is {\it not} invariant with respect 
to conformal rescalings. Later these two generalizations were shown 
to be special cases of a more general construction based on the bundle 
of spinors with $k$ unprimed and $l$ primed indices [14], and they can 
be recovered from that corresponding to the $(k,l)=(1,0)$ spinor 
representation and its complex conjugate. Thus it is enough to consider 
the Chern--Simons functional constructed in the basic spinor 
representation.

The present paper is addressed to the problem of dynamics of vacuum 
general relativity using certain three dimensional (conformal) 
geometries and the Chern--Simons functional, both in the spacetime 
and the phase space, but from a slightly different point of view. We 
consider a special spinorial 3-geometry, built from the initial data 
(and not only from the spatial 3-metric), and the Chern--Simons 
functional $Y$ constructed in the basic spinor representation. We 
are arguing that the proper interpretation of the Misner and York 
times and the imaginary part of the spinor Chern--Simons functional 
is not `internal time', rather they are observables on the phase space 
by means of which the constraints of vacuum general relativity can be 
rewritten into a new form. First we show that this Chern--Simons 
functional is invariant with respect to infinitesimal spacetime 
conformal rescalings if and only if the vacuum Einstein equations 
are satisfied. Then we reformulate this result in the ADM phase space 
of vacuum general relativity, and we show that this Chern--Simons 
functional generates the time evolution of vacuum general relativity 
in the sense that the Hamiltonian constraint is just the condition 
that the Poisson bracket of the imaginary part of $Y$ and Misner's 
time be zero. (The Poisson bracket of ${\rm Re}\,Y$ and Misner's time 
is automatically zero.) The momentum constraint will also be 
reformulated as a condition on the Poisson bracket of ${\rm Im}\,Y$ 
and the integral of York's local time. 

In Section 2. we introduce the Chern--Simons functional $Y$ in a direct 
way, without referring to the Lorentzian vector bundle of [18,14], and 
review its most important properties that we need in what 
follows. In particular, we clarify the conformal properties of $Y$ and 
show that it is invariant with respect to infinitesimal spacetime 
conformal rescalings iff the constraint parts of the vacuum Einstein 
equations are satisfied. In Section 3. the base manifold $\Sigma$ is 
considered to be a spacelike hypersurface in the spacetime, and $Y$ is 
shown to be invariant with respect to infinitesimal spacetime conformal 
rescalings on {\it every} spacelike hypersurface iff the vacuum Einstein 
equations (and not only their constraint parts) are satisfied. Then, in 
Section 4, the structures on the ADM phase space $\Gamma_{ADM}$ that we 
need are reviewed, and, in particular, we introduce the Misner and York 
times and the spinor Chern--Simons functional as functions on $\Gamma
_{ADM}$. In Section 5. the notion of conformal rescalings is implemented 
in $\Gamma_{ADM}$, and it is shown how the vacuum constraints of 
Einstein's theory can be reformulated by means of ${\rm Im}\,Y$. 
Finally, in Section 6, we discuss the Chern--Simons functional on the 
Ashtekar phase space.

Our conventions are mostly those of [19] (and follow [18,14]): 
In particular, the exterior product is defined as the anti-symmetrized 
tensor product and the spacetime signature is --2. The curvature and 
Ricci tensors e.g. of the covariant derivative operator $\nabla_a$ are 
defined by $-{}^4R^a{}_{bcd}W^bX^cY^d:=\nabla_X\nabla_YW^a-\nabla_Y
\nabla_XW^a-\nabla_{[X,Y]}W^a$ and ${}^4R_{bd}:={}^4R^a{}_{bad}$, 
respectively, and the scalar curvature is ${}^4R:={}^4R_{ab}g^{ab}$. 
Thus, in the presence of matter, Einstein's equations take the form 
${}^4G_{ab}:={}^4R_{ab}-{1\over2}{}^4Rg_{ab}=-\kappa T_{ab}$, where 
$\kappa:=8\pi G$ and $G$ is Newton's gravitational constant. Our 
differential geometric background is based primarily on [20]. 
\bigskip
\bigskip

\bigskip
\ni
{\lbf 2. The spinor Chern--Simons functional of the triples $(\Sigma,
q_{\mu\nu},\chi_{\mu\nu})$}\par
\bigskip
\ni
Let $\Sigma$ be a connected, orientable 3-manifold, ${\bf S}^A(\Sigma)$ 
a complex vector bundle of rank two over $\Sigma$ and $\bar{\bf S}^{A'}
(\Sigma)$ its complex conjugate bundle. Let $\varepsilon_{AB}$ be a 
symplectic and $t_{AA'}$ a positive definite Hermitian fibre metric on 
${\bf S}^A(\Sigma)$. We assume that $\varepsilon_{AB}$ and $t_{AA'}$ 
are compatible in the sense that $2\varepsilon^{A'B'}t_{AA'}t_{BB'}=
\varepsilon_{AB}$, and hence $\varepsilon^{AB}\varepsilon^{A'B'}t_{BB'}$ 
is just the inverse $t^{AA'}$ of $t_{AA'}$, where the inverses are 
defined by $\varepsilon^{AC}\varepsilon_{BC}:=\delta^A_B$ and $2t^{AA'}
t_{BA'}:=\delta^A_B$, respectively. We identify ${\bf S}^A(\Sigma)$ 
with its dual ${\bf S}_A(\Sigma)$ via $\varepsilon_{AB}$, and $\bar
{\bf S}^{A'}(\Sigma)$ with $\bar{\bf S}_{A'}(\Sigma)$ via $\varepsilon
_{A'B'}$. An element $X^{AA'}$ of ${\bf S}^A(\Sigma)\otimes\bar{\bf S}
^{A'}(\Sigma)$ is called real if $\bar X^{AA'}=X^{AA'}$, and note that 
$\varepsilon_{AB}\varepsilon_{A'B'}$ is a Lorentzian fibre metric with 
signature $(+,-,-,-)$ on the subbundle of the real elements of ${\bf S}
^A(\Sigma)\otimes\bar{\bf S}^{A'}(\Sigma)$. $t_{AA'}$ can also be 
considered as a (real) element of ${\bf S}_A(\Sigma)\otimes\bar{\bf S}
_{A'}(\Sigma)$, and it is timelike with unit length. $P^{AA'}_{BB'}:=
\delta^A_B\delta^{A'}_{B'}-t^{AA'}t_{BB'}={1\over2}(\delta^A_B\delta
^{A'}_{B'}-2t^A{}_{B'}t^{A'}{}_B)$ is the projection to the subbundle 
of the elements of ${\bf S}^A(\Sigma)\otimes\bar{\bf S}^{A'}(\Sigma)$ 
orthogonal to $t_{AA'}$, and hence any section $K^{AA'}$ of ${\bf S}^A
(\Sigma)\otimes\bar{\bf S}^{A'}(\Sigma)$ has a unique decomposition 
into the sum of a section proportional to and orthogonal to $t^{AA'}$ 
as $K^{AA'}=Nt^{AA'}+N^{AA'}$. By a theorem of Stiefel every orientable 
3-manifold is parallelizable, i.e. its tangent bundle is trivial (see 
e.g. [21]), thus if ${\bf S}^A(\Sigma)$ is chosen to be trivial then 
there is a globally defined (base point preserving) bundle injection 
$\Theta:T\Sigma\rightarrow{\bf S}^A(\Sigma)\otimes\bar{\bf S}^{A'}
(\Sigma)\,:X^\mu\mapsto X^{AA'}:=X^\mu\Theta^{AA'}_\mu$, which is an 
isomorphism between the tangent bundle $T\Sigma$ of $\Sigma$ and the 
bundle of the real elements of ${\bf S}^A(\Sigma)\otimes\bar{\bf S}
^{A'}(\Sigma)$ orthogonal to $t_{AA'}$. By $\Theta^{AA'}_\mu t_{AA'}
=0$ one has $\Theta^{AB}_\mu:=-\sqrt{2}\Theta^{AA'}_\mu t_{A'}{}^B=
\Theta^{(AB)}_\mu$, and the pull back $q_{\mu\nu}:=\Theta^{AA'}_\mu
\Theta^{BB'}_\nu\varepsilon_{AB}\varepsilon_{A'B'}$ of the Lorentzian 
fibre metric to $T\Sigma$ along $\Theta$ is real and negative definite. 
Thus the embedding $\Theta^{AA'}_\mu$ defines an $SU(2)$-spinor 
structure on $\Sigma$ with $SU(2)$ soldering form $\Theta^{AB}_\mu$, 
which therefore satisfies $\Theta^A_{\mu B}\Theta^B_{\nu C}={{\rm i}
\over\sqrt2}\varepsilon_{\mu\nu}{}^\rho\Theta^A_{\rho C}-{1\over2}
q_{\mu\nu}\delta^A_C$. Here $\varepsilon_{\mu\nu\rho}$ is the metric 
volume 3-form on $\Sigma$ determined by $q_{\mu\nu}$. 
Although there might be other $SU(2)$-spinor structures on $\Sigma$, 
labeled by the elements of the cohomology group $H^1(\Sigma,{\bf Z}
_2)$, by spinors on $\Sigma$ we mean the elements of the trivial bundle 
${\bf S}^A(\Sigma)$ above. By the triviality ${\bf S}^A(\Sigma)$ 
admits globally defined spin frame fields $\{\varepsilon^A_{\uA}\}$, 
${\uA}=0,1$, normalized by $\varepsilon_{AB}\varepsilon^A_{\uA}
\varepsilon^B_{\uB}=\epsilon_{\uA\uB}$, where $\epsilon_{\uA\uB}$ is 
the alternating Levi-Civita symbol. The dual spin frame field will be 
denoted by $\{\varepsilon^{\uA}_A\}$. If the basis $\{\varepsilon^A
_{\uA}\}$ is normalized with respect to $t_{AA'}$, too, i.e. $t_{AA'}
\varepsilon^A_{\uA}\bar\varepsilon^{A'}_{{\uA}'}=\sigma^0_{\uA{\uA}'}
:={1\over\sqrt{2}}{\rm diag}(1,1)$, then it is called an $SU(2)$-spin 
frame. Let $\sigma^{\ua}_{{\uA}{\uA}'}=(\sigma^0_{{\uA}{\uA}'},\sigma
^{\bi}_{{\uA}{\uA}'})$, ${\ua}=0,...,3$ and ${\bi}=1,2,3$, be the 
standard $SL(2,{\bf C})$ Pauli matrices (including the factor $1/
\sqrt2$), and define the three 1-forms $\vartheta^{\bi}_\mu:=\Theta
^{AA'}_\mu\varepsilon^{\uA}_A\bar\varepsilon^{{\uA}'}_{A'}\sigma^{\bi}
_{{\uA}{\uA}'}$. It is easy to see that $\{\vartheta^{\bi}_\mu\}$ is 
$q_{\mu\nu}$-orthonormal if $\{\varepsilon^A_{\uA}\}$ is an 
$SU(2)$-spin frame: $\vartheta^{\bi}_\mu\vartheta^{\bj}_\nu\eta_{\bi
\bj}=q_{\mu\nu}$, where $\eta_{\bi\bj}:={\rm diag}(-1,-1,-1)$. 
Furthermore, if $\{\varepsilon^A_{\uA}\}$ is a global spin frame field 
in ${\bf S}^A(\Sigma)$, then $\{\vartheta^{\bi}_\mu\}$ is a global 
1-form frame field in $T^*\Sigma$.

The metric $q_{\mu\nu}$ on $T\Sigma$ defines the Levi-Civita 
covariant derivative $D_\mu$, whose action can be extended to ${\bf 
S}^A(\Sigma)$ by requiring $D_\mu\Theta^{AA'}_\nu=0$, $D_\mu
\varepsilon_{AB}=0$ and $D_\mu t_{AA'}=0$. If $\{\varepsilon^A_{\uA}
\}$ is an $SU(2)$-spin frame, then the connection 1-form for $D_\mu$ 
on ${\bf S}^A(\Sigma)$ is well known to be expressible by the Ricci 
rotation coefficients defined in the basis $\{\vartheta^{\bi}_\mu\}$. 
Explicitly, $\gamma^{\uA}_{\mu{\uB}}:=\varepsilon^{\uA}_AD_\mu
\varepsilon^A_{\uB}=-{{\rm i}\over2\sqrt{2}}(\vartheta^{\bi}_\nu D_\mu 
e^\nu_{\bj})\varepsilon_{\bi}{}^{\bj\bk}\sigma^{\uA}_{{\bk}{\uB}}$, 
where $\{e^\mu_{\bi}\}$ is the vector basis in $T\Sigma$ dual to $\{
\vartheta^{\bi}_\mu\}$, $\varepsilon_{\bi\bj\bk}$ is the alternating 
Levi-Civita symbol, and $\sigma^{\uA}_{{\bi}{\uB}}:=\sqrt{2}\sigma
^{\uA}_{{\bi}{\uB}'}\epsilon^{{\uB}'{\uC}'}\sigma^0_{\uB{\uC}'}$, 
the standard $SU(2)$ Pauli matrices (including the coefficient $1/
\sqrt2$). (Boldface indices are moved by $\eta_{\bi\bj}$ and its 
inverse. Thus $\varepsilon^{\bi\bj\bk}$ is {\it minus} the Levi-Civita 
symbol $\epsilon^{\bi\bj\bk}$.) Next we define another covariant 
derivative ${\cal D}_\mu$ on the spinor bundle ${\bf S}^A(\Sigma)$ by 
requiring that 
i. ${\cal D}_\mu\varepsilon_{AB}=0$, 
ii. $\chi_{\mu\nu}:=\chi_{\mu AA'}\Theta^{AA'}_\nu:=({\cal D}_\mu t
   _{AA'})\Theta^{AA'}_\nu=\chi_{(\mu\nu)}$, and 
iii. ${\cal D}_\mu(v^\nu\Theta^{BB'}_\nu)P^{AA'}_{BB'}=(D_\mu 
   v^\nu)\Theta^{AA'}_\nu$ for any vector field $v^\nu$ on $\Sigma$. 
Since ${\cal D}_\mu t^{AA'}$ and ${\cal D}_\mu(e^\nu_{\bi}\Theta^{A
A'}_\nu)$ are the ${\cal D}_\mu$-derivative of pointwise independent 
cross sections of ${\bf S}^A(\Sigma)\otimes\bar{\bf S}^{A'}(\Sigma)$, 
where $\{e^\nu_{\bi}\}$ is any $q_{\mu\nu}$-orthonormal frame field, 
for given $\Theta^{AA'}_\mu$, $q_{\mu\nu}$ and $\chi_{\mu\nu}$ the 
conditions i.--iii. completely fix the derivative ${\cal D}_\mu$ on 
it. But since ${\cal D}_\mu$ annihilates not only the Lorentzian 
metric $\varepsilon_{AB}\varepsilon_{A'B'}$ on ${\bf S}^A(\Sigma)
\otimes\bar{\bf S}^{A'}(\Sigma)$ but $\varepsilon_{AB}$ itself, 
${\cal D}_\mu$ is completely determined on ${\bf S}^A(\Sigma)$, too. 
Geometrically, $\chi_{\mu}{}^{AA'}$ measures the 
non-$t_{AA'}$-metricity of ${\cal D}_\mu$. To compare the derivatives 
${\cal D}_\mu$ and $D_\mu$, first let us observe that $\chi_\mu{}^{AB}
:=-\sqrt{2}({\cal D}_\mu t^{AA'})t_{A'}{}^B=\chi_\mu{}^{(AB)}$, and 
consider the ${\cal D}_\mu$-derivative 
of $K^{AA'}=Nt^{AA'}+N^{AA'}$. It is ${\cal D}_\mu K^{AA'}=(D_\mu N)
t^{AA'}+N{\cal D}_\mu t^{AA'}+{\cal D}_\mu N^{AA'}=D_\mu K^{AA'}+(\chi
_\mu{}^{AA'}t_{BB'}-t^{AA'}\chi_{\mu BB'})K^{BB'}$. Next, choosing 
$K^{AA'}$ to be $\lambda^A\bar\mu^{A'}$, this yields ${\cal D}_\mu
\lambda^A=D_\mu\lambda^A-{1\over\sqrt{2}}\chi_\mu{}^A{}_B\lambda^B$. 
Then the curvature of ${\cal D}_\mu$ can be computed easily: $F^A{}
_{B\mu\nu}=R^A{}_{B\mu\nu}+{1\over\sqrt{2}}(D_\mu\chi_\nu{}^A{}_B-D
_\nu\chi_\mu{}^A{}_B)-{1\over2}(\chi_\mu{}^A{}_C\chi_\nu{}^C{}_B-\chi
_\nu{}^A{}_C\chi_\mu{}^C{}_B)$, where $R^A{}_{B\mu\nu}$ is the 
curvature of $D_\mu$ on ${\bf S}^A(\Sigma)$. The connection 1-form 
and curvature 2-form of ${\cal D}_\mu$ in the global spin frame field 
above are $\Gamma^{\uA}_{\mu{\uB}}:=\varepsilon^{\uA}_A{\cal D}_\mu
\varepsilon^A_{\uB}$ and $F^{\uA}{}_{{\uB}\mu\nu}:=\varepsilon^{\uA}_A
\varepsilon^B_{\uB}F^A{}_{B\mu\nu}$, respectively. Since they are 
globally defined on $\Sigma$, we can form the Chern--Simons functional 

$$
Y[\Gamma^{\uA}{}_{\uB}]:=\int_\Sigma\Bigl(F^{\uA}{}_{{\uB}\mu\nu}
\Gamma^{\uB}_{\rho{\uA}}+{2\over3}\Gamma^{\uA}_{\mu{\uB}}\Gamma^{\uB}
_{\nu{\uC}}\Gamma^{\uC}_{\rho{\uA}}\Bigr){1\over3!}\delta^{\mu\nu\rho}
_{\alpha\beta\gamma}, \eqno(2.1)
$$
\ni
provided the integral exists, e.g. if $\Sigma$ is closed or if $(
\Sigma,q_{\mu\nu},\chi_{\mu\nu})$ is asymptotically flat in the sense 
that both $q_{\mu\nu}$ and $r\chi_{\mu\nu}$ tend to a flat metric and 
zero, respectively, at least logarithmically with the radial distance 
$r$ [14]. $Y[\Gamma^{\uA}{}_{\uB}]$ is known to be invariant with 
respect to orientation preserving diffeomorphisms of $\Sigma$ onto 
itself, and also with respect to basis transformations $\Lambda:\Sigma
\rightarrow SL(2,{\bf C})$ of the spin frame that are homotopic to the 
identity transformation (``small gauge transformations"). However, 
under large (i.e. not small) gauge transformations $Y[\Gamma^{\uA}{}
_{\uB}]$ changes by $16\pi^2N$, where $N$ is integer and depends only 
on the homotopy class of $\Lambda$. In particular, the imaginary part 
of $Y[\Gamma^{\uA}{}_{\uB}]$ is gauge invariant. Since the quotient 
$SL(2,{\bf C})/SU(2)$ is homeomorphic to ${\bf R}^3$, furthermore any 
two continuous mappings $\lambda:\Sigma\rightarrow{\bf R}^3$ are 
homotopic, any global spin frame field can be transformed into an 
$SU(2)$-spin frame field by a small gauge transformation. Therefore, 
$Y[\Gamma^{\uA}{}_{\uB}]$ can always be computed in an $SU(2)$-spin 
frame, and hence it is determined completely by $\vartheta^{\bi}_\mu$ 
and $\chi_{\mu\nu}$. Furthermore, since $Y[\Gamma^{\uA}{}_{\uB}]$ 
modulo $16\pi^2$ is gauge invariant, it is a functional only of $q
_{\mu\nu}$ and $\chi_{\mu\nu}$ and will be denoted by $Y[q_{\mu\nu},
\chi_{\mu\nu}]$, or simply by $Y$. Its variational derivatives are 
 
$$\eqalignno{
{\delta Y\over\delta \chi_{\mu\nu}}=&2\sqrt{\vert q\vert}H^{\mu\nu}+
   2{\rm i}\sqrt{\vert q\vert}\Bigl(R^{\mu\nu}-{1\over2}Rq^{\mu\nu}+V
   ^{\mu\nu}-{1\over2}Vq^{\mu\nu}\Bigr),&(2.2.a) \cr
{\delta Y\over\delta q_{\mu\nu}}=&-\sqrt{\vert q\vert}\Bigl(B^{\mu
   \nu}+\chi^{(\mu}{}_\rho H^{\nu)\rho}\Bigr)-{\rm i}\sqrt{\vert q
   \vert}\Bigl(\varepsilon^{\rho\omega(\mu}D_\rho H^{\nu)}{}_\omega+
   {1\over2}D^{(\mu}\bigl(D_\rho\chi^{\nu)\rho}-D^{\nu)}\chi\bigr)-\cr
&\hskip 12pt -{1\over2}q^{\mu\nu}D_\omega\bigl(D_\rho\chi^{\rho\omega}
   -D^\omega\chi\bigr)+\chi^{(\mu}{}_\rho\bigl(R^{\nu)\rho}-{1\over2}
   q^{\nu)\rho}R+V^{\nu)\rho}-{1\over2}q^{\nu)\rho}V\bigr)\Bigr), 
   &(2.2.b)\cr}
$$
\ni
where $R_{\mu\nu}$ is the Ricci tensor of the Riemannian 3-geometry 
$(\Sigma,q_{\mu\nu})$, $R$ is its curvature scalar, we used the 
notations $V_{\mu\nu}:=\chi\chi_{\mu\nu}-\chi_{\mu\rho}\chi^\rho{}
_\nu$ and $V:=V^\mu{}_\mu=\chi^2-\chi^{\mu\nu}\chi_{\mu\nu}$, and 
introduced the tensor fields $H_{\mu\nu}:=-\varepsilon_{\rho\omega
(\mu}D^\rho\chi^\omega{}_{\nu)}$ and $B_{\mu\nu}:=-\varepsilon_{\rho
\omega(\mu}D^\rho R^\omega{}_{\nu)}-\varepsilon_{\rho\omega(\mu}D
^\rho V^\omega{}_{\nu)}+{1\over2}\chi^\rho{}_{(\mu}\varepsilon_{\nu)
\rho\omega}D_\alpha(\chi^{\alpha\omega}-\chi q^{\alpha\omega})$. 
Note that both $H_{\mu\nu}$ and $B_{\mu\nu}$ are traceless and 
symmetric, and for vanishing $\chi_{\mu\nu}$ the latter reduces to 
the Cotton--York tensor of $(\Sigma,q_{\mu\nu})$. We have shown that 
the stationary points of $Y[q_{\mu\nu},\chi_{\mu\nu}]$ are precisely 
those triples $(\Sigma,q_{\mu\nu},\chi_{\mu\nu})$ that can be locally 
isometrically embedded into the Minkowski spacetime with first and 
second fundamental forms $q_{\mu\nu}$ and $\chi_{\mu\nu}$, 
respectively [18].

Let $\Omega:\Sigma\rightarrow(0,\infty)$, $\dot\Omega:\Sigma
\rightarrow{\bf R}$ be smooth functions. The conformal rescaling of 
the metrics $\varepsilon_{AB}$ and $t_{AA'}$ and the connection ${\cal 
D}_\mu$ by the pair $(\Omega,\dot\Omega)$ is defined by $\varepsilon
_{AB}\mapsto\Omega\varepsilon_{AB}$, $t_{AA'}\mapsto\Omega t_{AA'}$ 
and $\chi_{\mu\nu}\mapsto\Omega\chi_{\mu\nu}+\dot\Omega q_{\mu\nu}$. 
This rescaling yields the change $\Gamma^{\uA}_{\mu{\uB}}\mapsto\tilde
\Gamma^{\uA}_{\mu{\uB}}:=\Gamma^{\uA}_{\mu{\uB}}-{1\over\sqrt2}\Omega
^{-1}(\delta^\rho_\mu\dot\Omega+{\rm i}\varepsilon_\mu{}^{\nu\rho}D
_\nu\Omega)\Theta^{\uA}_{\rho{\uB}}$ of the connection 1-form, and, by 
a straightforward calculation, the following change of $Y[\Gamma^{\uA}
{}_{\uB}]$:  

$$\eqalign{
Y[\tilde\Gamma^{\uA}{}_{\uB}]&=Y[\Gamma^{\uA}{}_{\uB}]-2{\rm i}\int
  _\Sigma\Bigl\{\Omega^{-1}\Bigl({1\over2}\bigl(R+V\bigr)\dot\Omega+
  D_\mu\bigl(\chi^{\mu\nu}-q^{\mu\nu}\chi\bigr)(D_\nu\Omega)\Bigr)+\cr
&+\Omega^{-2}\Bigl(\chi\dot\Omega^2+2\dot\Omega q^{\mu\nu}D_\mu(D_\nu
  \Omega)+\chi^{\mu\nu}(D_\mu\Omega) \,(D_\nu\Omega)\Bigr)+\Omega^{-3}
  \Bigl(\dot\Omega^3-\dot\Omega q^{\mu\nu}(D_\mu\Omega)\,(D_\nu\Omega)
  \Bigr)\Bigr\}\,{\rm d}\Sigma, \cr}\eqno(2.3)
$$
\ni
where ${\rm d}\Sigma:={1\over3!}\varepsilon_{\mu\nu\rho}$, the metric 
volume element. In particular, the real part of $Y[\Gamma^{\uA}{}
_{\uB}]$ is invariant with respect to conformal rescalings. Let 
$(\Omega(u),\dot\Omega(u))$ be a smooth 1-parameter family of conformal 
factors such that $\Omega(0)=1$ and $\dot\Omega(0)=0$ (i.e. they 
represent the identity transformation at $u=0$), and define $\delta
\Omega:=({\rm d}\Omega(u)/{\rm d}u)_{u=0}$ and $\delta\dot\Omega:=
({\rm d}\dot\Omega(u)/{\rm d}u)_{u=0}$. By (2.3) under such an 
infinitesimal conformal rescaling $Y[\Gamma^{\uA}{}_{\uB}]$ transforms 
as $\delta Y[\Gamma^{\uA}{}_{\uB}]=-2{\rm i}\int_\Sigma({1\over2}(R+V)
\delta\dot\Omega+D_\mu(\chi^{\mu\nu}-q^{\mu\nu}\chi)D_\nu\delta\Omega)
\sqrt{\vert q\vert}{\rm d}^3x$. Remarkably enough, it is just the 
constraint parts of the spacetime Einstein tensor that appear in 
$\delta Y$ in spite of the fact that we have not assumed anything 
about the field equations, and {\it the spinor Chern--Simons functional 
is invariant with respect to infinitesimal conformal rescalings iff 
$R+\chi^2-\chi_{\mu\nu}\chi^{\mu\nu}=0$ and $D_\mu(\chi^\mu{}_\nu-
\delta^\mu_\nu\chi)=0$}.

The connection ${\cal D}_\mu$ on ${\bf S}^A(\Sigma)$ determines 
a unique connection on the bundle ${\bf S}^{(A_1...A_k)(B'_1...B'_l)}
(\Sigma)$ of totally symmetric spinors on $\Sigma$ with $k$ unprimed 
and $l$ primed indices. The Chern--Simons functional built from this 
connection, denoted by $Y_{(k,l)}$, has been shown to be determined 
completely by the spinor Chern--Simons functional and its complex 
conjugate [14]: $Y_{(k,l)}={1\over6}(k+1)(l+1)(k(k+2)Y[\Gamma^{\uA}{}
_{\uB}]+l(l+2)\overline{Y[\Gamma^{\uA}{}_{\uB}]})$. In particular, the 
Chern--Simons functional in the $(k,k)$ (real tensor) representations 
is proportional to the real part of $Y[\Gamma^{\uA}{}_{\uB}]$. The 
stationary points of $Y_{(k,k)}$, characterized by $H_{\mu\nu}=0$ and 
$B_{\mu\nu}=0$, are precisely those triples $(\Sigma,q_{\mu\nu},\chi
_{\mu\nu})$ that can be locally isometrically embedded into the 
conformal Minkowski spacetime with the first and second fundamental 
forms $q_{\mu\nu}$ and $\chi_{\mu\nu}$, respectively [18]. Furthermore, 
by (2.3) $Y_{(k,k)}$ is invariant with respect to conformal rescalings. 
In fact, $Y_{(k,k)}$ can be rewritten as the Chern--Simons functional 
built from the 3-surface local twistor connection on $\Sigma$, which 
is a manifestly conformally invariant expression [14]. Thus, in 
complete agreement with (2.3), it is the imaginary part of $Y[\Gamma
^{\uA}{}_{\uB}]$ that in general breaks the conformal invariance. 
\bigskip
\bigskip

\ni
{\lbf 3. The spinor Chern--Simons functional of the spacelike 
hypersurfaces}\par
\bigskip
\ni
Let $(M,g_{ab})$ be a spacetime, $\theta_t:\Sigma\rightarrow M$, $t\in
{\bf R}$, a foliation of $M$ by spacelike hypersurfaces $\Sigma_t:=
\theta_t(\Sigma)$, and let $t_a$ be the future directed unit timelike 
normal to the leaves $\Sigma_t$ and $P^a_b:=\delta^a_b-t^at_b$ the 
corresponding projection to $\Sigma_t$. Since $M$ is diffeomorphic to 
$\Sigma\times{\bf R}$, $(M,g_{ab})$ admits a spinor structure, and the 
spinor structures on $(M,g_{ab})$ are in a 1--1 correspondence with 
those on $\Sigma$. Thus there is one spinor bundle ${\bf S}^A(M)$ over 
$M$ whose pull back to $\Sigma$ is just the trivial ${\bf S}^A
(\Sigma)$ above, and hence this ${\bf S}^A(M)$ is also globally 
trivializable. Let the corresponding $SL(2,{\bf C})$-soldering form 
on $M$ be $\vartheta^{AA'}_a$. Then the bundle embedding $\Theta^{AA'}
_\mu$ of the previous section can be identified with the composition 
of the differential $\theta_{t*}{}^a_\mu$ of the embedding $\theta_t$ 
and of the spacetime soldering form $\vartheta^{AA'}_a$. In particular, 
the spinor form $t^{AA'}:=t^a\vartheta^{AA'}_a$ of the normal is just 
the positive definite Hermitian fibre metric, and the pull back $\theta
_{t*}{}^a_\mu{\cal D}_a$ of the Sen connection ${\cal D}_a:=P^b_a\nabla
_b$ is just the covariant derivative ${\cal D}_\mu$ on ${\bf S}^A
(\Sigma)$. $q_{\mu\nu}$ becomes the pull back to $\Sigma$ of the 
induced metric $q_{ab}:=P^c_aP^d_bg_{cd}$ on $\Sigma_t$, $\varepsilon
_{\mu\nu\rho}$ the pull back of the induced volume form $\varepsilon
_{abc}:=\varepsilon_{abcd}t^d$, and $\chi_{\mu\nu}$ the pull back of 
the extrinsic curvature $\chi_{ab}:=P^c_aP^d_b\nabla_ct_d$ of $\Sigma
_t$ in $M$ with trace $\chi:=\chi_{ab}q^{ab}$. 
(Our choice $\varepsilon_{abc}:=\varepsilon_{abcd}t^d$ for the 
relation between the three and four dimensional volume forms is 
connected with the sign convention in the definition $\bar{\bf S}^{A'}
(\Sigma)\rightarrow{\bf S}^X(\Sigma):\bar\lambda^{A'}\mapsto-\sqrt{2}
\bar\lambda^{A'}t_{A'}{}^X$ for the primed-unprimed correspondence of 
the contravariant spinor indices: The spinor form of the intrinsic 
volume 3-form is $\varepsilon_{\mu\nu\rho}\Theta^\mu_{AX}\Theta^\nu
_{BY}\Theta^\rho_{CZ}=({\rm i}/\sqrt{2})(\varepsilon_{A(B}\varepsilon
_{Y)(C}\varepsilon_{Z)X}+\varepsilon_{X(B}\varepsilon_{Y)(C}
\varepsilon_{Z)A})$, which coincides with the unitary spinor form of 
the induced volume 3-form only if the latter is defined by the 
convention above. Thus, although in differential geometry the volume 
form of a hypersurface is defined by the contraction of the normal 
with the {\it first} index of the volume form of the embedding 
geometry, we adopt the standard sign conventions in the theory of 
spinors rather than the standard sign conventions in differential 
geometry.) 
If ${}^4G_{ab}$ and ${}^4C_{abcd}$ are the spacetime Einstein and Weyl 
tensors, respectively, then ${}^4G_{ab}t^at^b=-{1\over2}(R+V)$, ${}^4
G_{ab}t^aP^b_c=-D_a(\chi^a{}_c-\delta^a_c\chi)$, and the conformal 
electric and magnetic curvatures are $E_{ab}:={}^4C_{acbd}t^bt^d=-(R
_{ab}+V_{ab})+{1\over3}q_{ab}(R+V)+{1\over2}P^c_aP^d_b({}^4G_{cd}-{1
\over3}q_{cd}q^{ef}\,{}^4G_{ef})$ and $H_{ab}:={1\over2}\varepsilon
_{ae}{}^{cd}\,{}^4C_{cdbf}t^et^f=-\varepsilon_{cd(a}D^c\chi^d{}_{b)}$, 
respectively. Thus although $H_{ab}$ can be expressed by the induced 
metric and extrinsic curvature, in general $E_{ab}$ cannot. The part 
of $E_{ab}$ that is determined by the geometry of $\Sigma_t$ is ${}_0E
_{ab}:=-(R_{ab}+V_{ab})+{1\over3}q_{ab}(R+V)$. The spacetime conformal 
rescaling $g_{ab}\mapsto\Omega^2g_{ab}$ by $\Omega:M\rightarrow(0,
\infty)$ induces the transformations $q_{ab}\mapsto\Omega^2q_{ab}$, 
$\chi_{ab}\mapsto\Omega\chi_{ab}+t^e(\nabla_e\Omega)q_{ab}$, which, 
identifying $\dot\Omega$ with $t^e\nabla_e\Omega$, justifies our 
definition for the conformal rescaling of $(\Sigma,\varepsilon_{AB},
t_{AA'},\chi_{\mu\nu})$ in the previous section.

If $\{\varepsilon^A_{\uA}\}$, $\{\varepsilon^{\uA}_A\}$ is a 
(globally defined) dual spin frame field in $M$, then the connection 
1-form ${}^{4}\Gamma^{\uA}_{e{\uB}}:=\varepsilon^{\uA}_A\nabla_e
\varepsilon^A_{\uB}$ and the curvature 2-form ${}^{4}R^{\uA}{}_{{\uB}
cd}:=\varepsilon^{\uA}_A\varepsilon^B_{\uB}{}^{4}R^A{}_{Bcd}$ are 
globally defined on $M$, and their pull back to $\Sigma$ are just the 
connection and curvature forms, $\Gamma^{\uA}_{\mu{\uB}}(t)$ and $F
^{\uA}{}_{{\uB}\mu\nu}(t)$, of the spinor connection on $\Sigma$, 
respectively. Thus the pull back to $\Sigma$ of the `spacetime' 
Chern--Simons 3-form ${}^{4}R^{\uA}{}_{{\uB}[ab}{}^{4}\Gamma^{\uB}_{c]
{\uA}}+{2\over3}{}^{4}\Gamma^{\uA}_{[a\vert{\uB}\vert}{}^{4}\Gamma
^{\uB}_{b\vert{\uC}\vert}{}^{4}\Gamma^{\uC}_{c]{\uA}}$ is just the 
Chern--Simons 3-form built from the spinor connection $\Gamma^{\uA}
_{\mu{\uB}}(t)$ on the bundle ${\bf S}^A(\Sigma)$. This `spacetime' 
picture makes the conformal behaviour of $Y[\Gamma^{\uA}{}_{\uB}]$ 
more transparent. In fact, if $\Upsilon_{AA'}:=\Omega^{-1}\nabla_{AA'}
\Omega=\Omega^{-1}(\dot\Omega t_{AA'}+\Theta^\mu_{AA'}D_\mu\Omega)$, 
then 

$$
Y[\tilde\Gamma^{\uA}{}_{\uB}]=Y[\Gamma^{\uA}{}_{\uB}]-2{\rm i}\int
_{\Sigma_t}\Bigl\{-t_a\,{}^4G^{ab}\Upsilon_b+{\rm i}\varepsilon^{abcd}
\Bigl(\bigl({\cal D}_{AA'}\Upsilon_{B'C}\bigr)\Upsilon_{C'B}+{2
\over3}\Upsilon_{A'B}\Upsilon_{B'C}\Upsilon_{C'A}\Bigr)t_{DD'}\Bigr\}
{\rm d}\Sigma_t.\eqno(3.1)
$$
\ni
Thus, as we saw in the previous section, the invariance of $Y[\Gamma
^{\uA}{}_{\uB}]$ with respect to infinitesimal conformal rescalings is 
equivalent to ${}^4G_{ab}t^b=0$ at arbitrary point $p$ of a given 
hypersurface $\Sigma_t$, but boosting $\Sigma_t$ slightly in three 
independent ways at $p$ and repeating the previous argument, for the 
invariance of $Y[\Gamma^{\uA}{}_{\uB}]$ we get ${}^4G_{ab}=0$. 
Therefore, {\it the spinor Chern--Simons functional is invariant with 
respect to infinitesimal conformal rescalings on every Cauchy surface 
iff the vacuum Einstein equations are satisfied}. Thus vacuum general 
relativity can be reformulated as an invariance requirement on the 
Chern--Simons functional built from a special spinor geometry on ${\bf 
S}^A(\Sigma)$ over the 3-manifold $\Sigma$. The obstruction to the 
invariance of $Y$ is the spacetime Einstein tensor.

In this spacetime context we can calculate the `time evolution' 
of $Y[\Gamma^{\uA}{}_{\uB}(t)]$, too. To do this first recall that the 
embedding $\theta_t:\Sigma\rightarrow M$ defines a congruence of curves 
on $M$ by assigning the curve $\theta(t):=\theta_t(p)$ to the point 
$\theta_0(p)\in M$. Let us denote its tangent vector field by $K^a$, 
and decompose it into the sum of its parts normal and tangential to 
$\Sigma_t$: $K^a=Nt^a+N^a$. Then, using the identity $\L_{\bf K}={\rm 
d}\circ\iota_{\bf K}+\iota_{\bf K}\circ{\rm d}$ for the Lie derivative 
on the `spacetime' Chern--Simons 3-form, for the time evolution we 
obtain 

$$\eqalign{
{{\rm d}\over{\rm d}t}Y[\Gamma^{\uA}{}_{\uB}(t)]&=\int_{\Sigma_t}
  {1\over2}\,{}^{4}R_{{\uA}{\uB}ab}\,{}^{4}R^{{\uA}{\uB}}{}_{cd}
  \varepsilon^{abcd}\,N\,{\rm d}\Sigma_t=\cr
&=\int_{\Sigma_t}\Bigl\{4E_{ab}H^{ab}-{\rm i}\Bigl(2\bigl(E_{ab}E
  ^{ab}-H_{ab}H^{ab}\bigr)-{1\over2}{}^{4}G_{ab}\,{}^{4}G^{ab}+{1
  \over6}\,{}^{4}R^2\Bigr)\Bigr\}N{\rm d}\Sigma_t,\cr}\eqno(3.2)
$$
\ni
where we used the expressions for $E_{ab}$ and $H_{ab}$ above and 
those for the `constraint parts' of the spacetime Einstein tensor 
by the three dimensional quantities. The real part of (3.2) is 
invariant with respect to spacetime conformal rescalings, thus the 
time derivative of the Chern--Simons functionals defined in the real 
tensor representations is also conformally invariant. In general 
neither the real nor the imaginary part has definite sign. Note that 
the shift vector does not appear on the right hand side of (3.2), 
showing the invariance of $Y$ with respect to spatial diffeomorphisms. 
In the rest of the present paper we identify $\Sigma$ with its image 
$\Sigma_t$, and hence use only the Latin indices. 
\bigskip
\bigskip

\ni
{\lbf 4. The spinor Chern--Simon functional on the ADM phase space}\par
\bigskip
\ni
The classical ADM phase space of vacuum general relativity (see, e.g. 
[1-3]), $\Gamma_{ADM}$, is the set of the pairs of fields $(q_{ab},
\tilde p^{ab})$ on a connected orientable 3-manifold $\Sigma$, where 
the configuration variables are the negative definite 3-metrics $q
_{ab}$ and the canonically conjugate momentum variables are the 
densitized symmetric tensor fields $\tilde p^{ab}$. Thus the canonical 
symplectic 2-form $\omega$ is defined by $2\omega({\cal X},{\cal X}')
:=\int_\Sigma((\delta\tilde p^{ab})(\delta'q_{ab})-(\delta'\tilde p
^{ab})(\delta q_{ab})){\rm d}^3x$ for any tangent vectors ${\cal X}=
(\delta q_{ab},\delta\tilde p^{ab})$ and ${\cal X}'=(\delta'q_{ab},
\delta'\tilde p^{ab})$. In terms of the Cauchy data in a local 
coordinate system $\tilde p^{ab}=-{1\over2\kappa}\sqrt{\vert q\vert}
(\chi^{ab}-\chi q^{ab})$, where $q:=\det(q_{ef})$, and $\chi_{ab}$ is 
the extrinsic curvature of $\Sigma$ in the spacetime. Then the 
Hamiltonian and the momentum constraints of the vacuum general 
relativity are 

$$\eqalignno{
\tilde{\cal C}&:=-{1\over2\kappa}\sqrt{\vert q\vert}\Bigl(R-{4\kappa
  ^2\over\vert q\vert}\bigl(\tilde p^{ab}\tilde p^{cd}q_{ac}q_{bd}-{1
  \over2}\bigl[\tilde p^{ab}q_{ab}\bigr]^2\bigr)\Bigr)=0,  &(4.1.a)\cr
\tilde{\cal C}_a&:=2q_{ab}D_c\tilde p^{bc}=0.&(4.1.b)\cr}
$$
\ni
The corresponding constraint functions on $\Gamma_{ADM}$ are defined 
by $C[N,N^a]:=\int_\Sigma(\tilde{\cal C}N+\tilde{\cal C}_aN^a){\rm d}
^3x$ for arbitrary function $N$ and vector field $N^a$ on $\Sigma$. 
If $\L_{\bf N}$ denotes the Lie derivative operator along the vector 
field $N^a$, then their functional derivatives are 

$$\eqalignno{
{\delta C[N,N^a]\over\delta q_{ab}}={1\over2\kappa}\sqrt{\vert q\vert}
  &\Bigl\{N\Bigl(R^{ab}-Rq^{ab}+{8\kappa^2\over\vert q\vert}\bigl(
  \tilde p^{ac}q_{cd}\tilde p^{bd}-{1\over2}q_{cd}\tilde p^{cd}\tilde 
  p^{ab}\bigr)\Bigr)+\cr
&+D^aD^bN-q^{ab}D_eD^eN\Bigr\}-{1\over2}N\tilde{\cal C}q^{ab}+\L_{\bf 
  N}\tilde p^{ab},&(4.2.a)\cr
{\delta C[N,N^a]\over\delta\tilde p{}^{ab}}={4\kappa\over\sqrt{\vert 
  q\vert}}N&\Bigl(\tilde p_{ab}-{1\over2}q_{ab}q_{cd}\tilde p{}^{cd}
  \Bigr)-\L_{\bf N}q_{ab}.&(4.2.b)\cr}
$$
\ni 
It has already been shown that in the asymptotically flat case (with 
the standard $1/r$ fall-off and even parity for the metric and $1/r
^2$ fall-off and odd parity for the canonical momenta) the vanishing 
of these functional derivatives and $C[N,N^a]=0$ together imply that 
$N=0$ and $N^a=0$ [22], i.e. $C[N,N^a]=0$ defines a non-degenerate 
`surface' in $\Gamma_{ADM}$. However, in the closed case $C[N,N^a]$ 
does have critical points even if $C[N,N^a]=0$. In fact, if $\L_{\bf N}
q_{ab}=0$ (e.g. when $N^a$ itself is vanishing), then by (4.2.b) 
$\tilde p^{ab}=0$, and then by the Hamiltonian constraint $R=0$. But 
by $\tilde p^{ab}=0$ (4.2.a) takes the form $NR^{ab}+D^aD^bN-q^{ab}(D_e
D^eN+{1\over2}NR)=0$, implying that $D_eD^eN=0$ and $NR^{ab}=-D^aD^bN$. 
But the first, together with the compactness of $\Sigma$, implies that 
$N={\rm const}$, and hence the second implies that $q_{ab}$ is flat. 
Therefore, for flat $q_{ab}$ the pair $(q_{ab},0)\in\Gamma_{ADM}$ is a 
critical point of the constraint function $C[N,0]$ for constant $N$. 
Since, however, the shift vector is a part of the spacetime diffeo 
gauge freedom, $C[N,N^a]$ is expected to have critical points 
representing the flat spacetime. This result is in complete agreement 
with the classical result [23] that the closed flat spacetimes are 
unstable in the sense that not all solutions of the linearized 
constraints correspond to nearby solutions of the constraint equations 
themselves.\footnote{*}{I am grateful to Niall \'O Murchadha for this 
remark and for pointing out reference [23].}

Recall that a vector field ${\cal X}$ on $\Gamma_{ADM}$ is called the 
Hamiltonian vector field of the function $\Phi:\Gamma_{ADM}\rightarrow
{\bf R}$ if $2\omega({\cal X},{\cal Y})+{\cal Y}(\Phi)=0$ for every 
vector ${\cal Y}$. The vector field ${\cal X}_\Phi:=(\delta\Phi/\delta
\tilde p^{ab},-\delta\Phi/\delta q_{ab})$ is a Hamiltonian vector 
field of $\Phi$, and, in fact, for finite dimensional symplectic 
manifolds the analogous expression follows from the definition of the 
Hamiltonian vector fields. In infinite dimensional phase spaces, 
however, the non-degeneracy of $\omega$ itself does not imply its 
invertability. $\Gamma_{ADM}$ would have to be endowed with a reflexive 
Banach manifold structure. Thus, in lack of additional assumptions 
on $\omega$, the definition of the Hamiltonian vector fields above 
does not imply this explicit expression for ${\cal X}_\Phi$ [24]. 
Thus we call ${\cal X}_\Phi$ the Hamiltonian vector field of $\Phi$ 
in the strong sense. In particular, the Hamiltonian vector field of 
$C[N,N^a]$ in the strong sense is ${\cal X}_{C[N,N^e]}=(\delta C[N,
N^e]/\delta\tilde p^{ab},-\delta C[N,N^e]/\delta q_{ab})$. The flow 
on $\Gamma_{ADM}$ corresponding to $C[0,N^a]$ is the system of 
equations $\dot q_{ab}=-\L_{\bf N}q_{ab}$, $\dot{\tilde p}{}^{ab}=
-\L_{\bf N}\tilde p^{ab}$; i.e. it is the natural lift of the vector 
field $-N^a$ to the phase space. Hence $C[0,N^a]$ generates a spatial 
diffeomorphism on $\Sigma$. The flow corresponding to $C[N,0]$ is 
just the system of evolution equations of the initial value formulation 
of the vacuum general relativity with vanishing shift vector. The 
Poisson bracket of the constraint functions is well known to be $\{C[N,
N^a],C[M,M^a]\}=C[\L_{\bf N}M-\L_{\bf M}N,[N,M]^a+MD^aN-ND^aM]$. Thus 
$N^a\mapsto C[0,N^a]$ defines a Lie algebra homomorphism of the Lie 
algebra of vector fields ${\rm Vect}(\Sigma)$ into the Poisson algebra 
of functions $C^\infty(\Gamma_{ADM},{\bf R})$. For spatially closed 
spacetimes, which we are concentrating on for the sake of simplicity, 
the Hamiltonian is just the constraint with arbitrary lapse and shift: 
$H[N,N^a]:=-C[N,N^a]$. Therefore, in the Hamiltonian the two 
constraints play different roles: while $C[N,0]$ generates the proper 
evolution of the states with respect to the coordinate time, i.e. the 
{\it dynamics}, $C[0,N^a]$ generates only a smooth {\it kinematical 
symmetry} of the theory, i.e. it ensures the invariance of the theory 
with respect to {\it spatial} diffeomorphisms $\Sigma\rightarrow
\Sigma$ that are homotopic to the identity mapping of $\Sigma$ onto 
itself (``small diffeomorphisms') [1-3]. (For a different interpretation 
of the constraints see [15].)

Next let us define $V[n]:=\int_\Sigma n\sqrt{\vert q\vert}{\rm d}^3x$ 
and $T[f]:={2\over3}\int_\Sigma fq_{ab} \tilde p^{ab}{\rm d}^3x$ for 
any fixed $n,f:\Sigma\rightarrow{\bf R}$. If $n$ is chosen to be the 
characteristic function of a subset $D\subset\Sigma$ then $V[n]$ 
becomes the metric volume ${\rm Vol}(D)$ of $D$, and $T[f]$ is the 
integral of York's local time smeared by $f$. (The area of a smooth 
orientable 2-surface ${\cal S}$ can also be recovered in a similar 
way.) Their Poisson bracket is $\{T[f],V[n]\}=V[fn]$, i.e. $T[1]$ acts 
on $V[n]$ as identity and hence $V[n]$ changes exponentially along the 
integral curves of the Hamiltonian vector fields of $T[1]$: If we write 
$V[n]=:V_0 \exp(v[n])$ for some constant $V_0$, then we have $\{T[1],
v[n]\}=1$. Misner's time is $-{1\over3}v[n]$. Neither $V[n]$ nor $T[f]$ 
has critical points on $\Gamma_{ADM}$. Their time evolution is: 

$$\eqalignno{
\dot V[n]&:=\Bigl\{H[N,N^a],V[n]\Bigr\}=-V[\L_{\bf N}n]+{3\kappa\over2}
  T[Nn], &(4.3.a)\cr
\dot T[f]&:=\Bigl\{H[N,N^a],T[f]\Bigr\}=-C[fN,0]-T[\L_{\bf N}f]-{2\over3
  \kappa}V[f(D_eD^eN+RN)].  &(4.3.b)\cr}
$$
\ni
Since in general $T[f]$ does not have any definite sign, $V[n]$, i.e. 
Misner's time, can in fact be monotonic only in rather special 
situations. Similarly, if $f$ is constant and the Hamiltonian 
constraint is satisfied, then $T[f]$ is monotonic only for those {\it 
specific} lapses $N$ for which the integral of $RN$ on $\Sigma$ is 
positive or negative. Such lapses always exist if $R$ is not 
identically vanishing, ensuring the monotonity of $T[f]$, at least in 
a small coordinate time interval.

Since the spinor Chern--Simons functional $Y[\Gamma^{\uA}{}_{\uB}]$ 
modulo $16\pi^2$ is a well defined function of $q_{ab}$ and $\chi_{ab}$, 
it defines a function $Y=Y[q_{ab},\tilde p^{ab}]$ on the ADM phase 
space as well. Note that $Y$ does not depend on any smearing function, 
furthermore, it is a dimensionless quantity. Its variational 
derivatives with respect to the ADM variables can be calculated using 
(2.2). They are 

$$\eqalignno{
{\delta Y\over\delta q_{ab}}=&\kappa\Bigl(-6\tilde p^{(a}{}_c\bigl(
   H^{b)c}-{\rm i}\,{}_0E^{b)c}\bigr)+2q^{ab}\tilde p^{cd}\bigl(H_{cd}
   -{\rm i}\,{}_0E_{cd}\bigr)+\tilde p\bigl(H^{ab}-{\rm i}\,{}_0E^{ab}
   \bigr)\Bigr)-\cr
&-{\rm i}\sqrt{\vert q\vert}\varepsilon^{cd(a}D_c\bigl(H^{b)}{}_d-{\rm 
   i}\,{}_0E^{b)}{}_d\bigr)+{{\rm i}\over2}\kappa\Bigl(D^{(a}\tilde
   {\cal C}^{b)}-q^{ab}D_c\tilde{\cal C}^c\Bigr)-{\kappa^2\over\sqrt{
   \vert q\vert}}\tilde p^{(a}{}_c\varepsilon^{b)cd}\tilde{\cal C}_d, 
   &(4.4.a)\cr
{\delta Y\over\delta\tilde p^{ab}}=&-4\kappa\Bigl(H_{ab}-{\rm i}{}_0
   E_{ab}\Bigr)+{2\over3}\kappa^2{{\rm i}\over\sqrt{\vert q\vert}}\,
   \tilde{\cal C}q_{ab}. &(4.4.b)\cr}
$$
\ni
By (4.4) the critical points of $Y$ are those for which ${}_0E_{ab}
=0$, $H_{ab}=0$, $\tilde{\cal C}=0$, $D_{(a}\tilde{\cal C}_{b)}=0$ and 
$\tilde p^{(a}{}_c\varepsilon^{b)cd}\tilde{\cal C}_d=0$. As we have 
already shown [18], ${}_0E_{ab}=0$, $H_{ab}=0$ and $D_{(a}\tilde{\cal 
C}_{b)}=0$ imply that $\tilde{\cal C}_a=0$. Thus, in particular, the 
critical points of $Y$ are all on the constraint surface, and, as we 
noted in Section 1, they represent initial data for locally flat 
spacetimes. The time evolution of $Y$, defined by $\dot Y:=\{H[N,N^a],
Y\}=-\{C[N,N^a],Y\}$, is just (3.2), where ${}^4G_{ab}=0$ (and hence 
${}^4R=0$ and $E_{ab}={}_0E_{ab}$). Thus, in particular, the Poisson 
bracket of $Y$ with the momentum constraint is zero, expressing its 
invariance with respect to small diffeomorphisms in the symplectic 
framework. However, apart from special configurations describing e.g. 
Petrov III. or N. spacetimes, the Poisson bracket with the Hamiltonian 
constraint is non-zero, and, apart from exceptional cases again, its 
sign is not definite. Thus the Hamiltonian vector field of neither 
${\rm Re}\,Y$ nor ${\rm Im}\,Y$ is tangent to the constraint surface 
in $\Gamma_{ADM}$, and neither ${\rm Re}\,Y$ nor ${\rm Im}\,Y$ is 
monotonic during the time evolution. On the other hand, for {\it 
specific} lapses ${\rm Im}\,\dot Y$ can be ensured to be positive, and 
hence, in a small coordinate time interval, ${\rm Im}\,Y$ can be used 
as an `internal time function'.

To summarize, these specific candidates for the `internal time 
function' have two main drawbacks: First, they are monotonic only for 
{\it specific} lapse functions instead of any $N$, and, second, they 
are not globally defined. In fact, the first implies the second: Since 
the lapse must be chosen to be `adapted' to the initial state $(q_{ab},
\tilde p^{ab})$ to ensure the positivity of the time derivative of the 
`time function', this `adaptation' may go wrong during the evolution 
of the state. Thus a good natural time variable would have to be a 
function $\tau:\Gamma_{ADM}\rightarrow{\bf R}$ whose derivative $\dot
\tau:=\{H[N,N^a],\tau\}$ is positive {\it for any positive lapse} $N$. 
Such a derivative could be, for example, $V[n]$ with some non-negative 
smearing function $n$, or the Bel--Robinson `energy' $E_{BR}[n]:=\int
_\Sigma({}_0E_{ab}\,{}_0E^{ab}+H_{ab}\,H^{ab})n\sqrt{\vert q\vert}{\rm 
d}^3x$ also with non-negative $n$. (The latter could be more natural 
because the vanishing of $E_{BR}[n]$ for any non-negative $n$, together 
with the vacuum constraints, implies that the corresponding initial 
data is flat, i.e. precisely the critical points of the vacuum 
constraints. It might be interesting to note that, apart from the sign 
in front of ${}_0E_{ab}\,{}_0E^{ab}$ in $E_{BR}[n]$, the imaginary 
part of (3.2) in vacuum is just this Bel--Robinson `energy'.) However, 
the question of the globality of $\tau$ would still be open, as it is 
not clear e.g. how the infinite dimensional versions of Poincare's 
recurrence theorem restrict the possibility of globally defined time 
functions, like in the phase space of mechanical systems. We will see 
in the next two sections that $V[n]$ plays, in fact, the role of 
determining the {\it scale} of a time parameter (rather than the time 
itself), but in a slightly different context. 

\bigskip
\bigskip

\ni
{\lbf 5. The role of spinor Chern--Simon functional in the dynamics 
of GR}\par
\bigskip
\ni
The conformal rescaling of $(q_{ab},\chi_{ab})$ yields the mapping 
$(q_{ab},\tilde p^{ab})\mapsto(\Omega^2q_{ab},\tilde p^{ab}+{1\over
\kappa}\Omega^{-1}\dot\Omega\sqrt{\vert q\vert}q^{ab})$ of the phase 
space onto itself. Thus the infinitesimal conformal rescaling, 
characterized by the pair of functions $(\delta\Omega,\delta\dot
\Omega)$, defines the vector field ${\cal K}:=(\delta q_{ab},\delta
\tilde p^{ab})=(2\delta\Omega q_{ab},{1\over\kappa}\delta\dot\Omega
\sqrt{\vert q\vert}q^{ab})$ on $\Gamma_{ADM}$. Then the effect of the 
infinitesimal conformal rescalings on a functionally differentiable 
function $F:\Gamma_{ADM}\rightarrow{\bf C}$ is ${\cal K}(F):=\int
_\Sigma({\delta F\over\delta q_{ab}}\delta q_{ab}+{\delta F\over\delta
\tilde p^{ab}}\delta\tilde p^{ab}){\rm d}^3x$. In particular, this 
action on the functions $V[n]$, $T[f]$ and $Y$, respectively, is 
${\cal K}(V[n])=3V[n\delta\Omega]$, ${\cal K}(T[f])=2T[f\delta\Omega]+
{2\over\kappa}V[f\delta\dot\Omega]$ and ${\cal K}(Y)=2{\rm i}\kappa\,
C[\delta\dot\Omega,D^a\delta\Omega]$. The latter is simply a 
reformulation of (the $u$-derivative of) (2.3) in the symplectic 
framework, and hence the vacuum constraints are equivalent to the 
invariance requirement ${\cal K}(Y)=0$ for any ${\cal K}$ above.

Next let us clarify whether ${\cal K}$ is a Hamiltonian vector field 
or not. Suppose that ${\cal K}$ is a Hamiltonian vector field of a 
function $\Phi$ in the strong sense, and hence $\Phi$ satisfies 

$$
{\delta\Phi\over\delta\tilde p^{ab}}=2q_{ab}\delta\Omega, \hskip 20pt
{\delta\Phi\over\delta q_{ab}}=-{1\over\kappa}\sqrt{\vert q\vert}q^{ab}
\delta\dot\Omega.\eqno(5.1)
$$
\ni
Then let us consider a smooth two-parameter family of points $(q_{ab}
(u,v),\tilde p^{ab}(u,v))$ of $\Gamma_{ADM}$, $u,v\in(-\epsilon,
\epsilon)$, and consider the function $\Phi$ as a function of the two 
parameters: $\Phi=\Phi(u,v)$. Using (5.1), the derivative of $\Phi(u,
v)$ with respect to $u$ at $u=0$, which is still a function of $v$, is 

$$\eqalign{
\delta_u\Phi:=\Bigl({{\rm d}\over{\rm d}u}\Phi(u,v)\Bigr)_{u=0}&=\int
  _\Sigma\Bigl\{{\delta\Phi\over\delta q_{ab}}\delta_uq_{ab}+{\delta\Phi
  \over\delta\tilde p^{ab}}\delta_u\tilde p^{ab}\Bigr\}{\rm d}^3x=\cr
&=\int_\Sigma\Bigl\{-{1\over\kappa}\delta\dot\Omega q^{ab}\sqrt{\vert 
  q\vert}\bigl(\delta_uq_{ab}\bigr)+2\delta\Omega q_{ab}\bigl(\delta_u
  \tilde p^{ab}\bigr)\Bigr\}{\rm d}^3x, \cr}
$$
\ni
and there is a similar expression for the $v$-derivative $\delta_v\Phi$, 
too. Since, however, $\Phi(u,v)$ is a smooth real valued function of 
two variables, the order of its $u$ and $v$ derivatives can be 
interchanged (`functional integrability condition'). Therefore, 

$$
0=\delta_v\delta_u\Phi-\delta_u\delta_v\Phi=2\int_\Sigma\delta\Omega
\Bigl(\bigl(\delta_u\tilde p^{ab}\bigr)\bigl(\delta_vq_{ab}\bigr)-
\bigl(\delta_v\tilde p^{ab}\bigr)\bigl(\delta_uq_{ab}\bigr)\Bigr){\rm 
d}^3x.
$$
\ni
However, apart from $\delta\Omega$, the right hand side is just the 
canonical symplectic 2-form evaluated on the vectors $(\delta_uq_{ab},
\delta_u\tilde p^{ab})$ and $(\delta_vq_{ab},\delta_v\tilde p^{ab})$, 
which is non-vanishing. Therefore, $\Phi$ can be a non-trivial 
solution of (5.1) only if $\delta\Omega=0$, and hence ${\cal K}$ can 
be a Hamiltonian vector field (in the strong sense) only for the 
special infinitesimal conformal rescalings with $\delta\Omega=0$. Thus 
the spatial conformal rescalings are not canonical transformations, 
but the temporal ones, characterized by $\delta\dot\Omega$, are. In 
this special case (5.1) can be integrated immediately: $\Phi=-{2\over
\kappa}V[\delta\dot\Omega]$. This and the expression for ${\cal K}
(Y)$ above lead to consider the Poisson bracket of $V[n]$ and $Y$. It 
is $\{Y,V[n]\}={\rm i}\kappa^2\,C[n,0]$, i.e. 

$$
\Bigl\{{\rm Re}\,Y,V[n]\Bigr\}=0, \hskip 22pt
\Bigl\{{\rm Im}\,Y,V[n]\Bigr\}=\kappa^2\,\,C[n,0].\eqno(5.2)
$$
\ni
Therefore, {\it the Hamiltonian constraint function $C[N,0]$ is a pure 
Poisson bracket of two naturally defined functions, and the geometric 
content of the Hamiltonian constraint is that the Hamiltonian vector 
field of the imaginary part of the spinor Chern--Simons functional be 
volume preserving, or, equivalently, ${\rm Im}\,Y$ must be constant 
along the flow of the Hamiltonian vector field of $V[N]$}. Thus, as 
Smolin and Soo have already 
realized, the imaginary part of $Y$ should be connected with the time 
evolution, but its proper interpretation is not an `intrinsic time 
function', rather it is a more elementary functional by means of which 
the constraint governing the dynamics of vacuum general relativity is 
generated by $V[N]$. Thus the time evolution is governed by the integral 
of a potential for a topological quantity (viz. the second Chern class 
of the spacetime) and the 3-volume. The lapse function enters $C[N,0]$ 
only through $V[N]$, while the Chern--Simons functional, being 
dimensionless and depending on no smearing function, appears to be 
some `universal master function'. $V[N]$ is not a time function either, 
rather it is a `sub-generator' of the dynamics of the vacuum general 
relativity, and, through the Poisson bracket with ${\rm Im}\,Y$, it 
determines the scale of the natural parameter of the integral curves 
of the Hamiltonian vector field of $C[N,0]$. The Hamiltonian vector 
field of the real part of $Y$ is automatically volume preserving, a 
manifestation of its conformal invariance in the symplectic framework. 
Remarkably enough, the 3-volume has already appeared in connection 
with the dynamics of general relativity: First, as Misner's time [11], 
or recently in the reduced Hamiltonian of Fischer and Moncrief [25] 
(see also [6,26]). By (5.2) it is perhaps more natural to interpret 
$C[N,0]=0$ as the condition that $Y$ must be constant along the flow 
of $V[N]$, because $V[N]$ does not have critical points.

Although for nonzero $\delta\Omega$ there is no differentiable 
function $\Phi$ on the phase space which would be a solution of (5.1) 
(i.e. there is no function $\Phi$ for which $2\omega({\cal K},{\cal X})
+{\cal X}(\Phi)=0$ would hold for {\it any} vector ${\cal X}$), for 
{\it specific} ${\cal X}$, namely for the Hamiltonian vector field 
${\cal X}_{{\rm Im}\,Y}$ of ${\rm Im}\,Y$, there may exist a function 
$\Phi$ for which $2\omega({\cal K},{\cal X}_{{\rm Im}\,Y})+{\cal X}
_{{\rm Im}\,Y}(\Phi)=0$ could hold. Or, in other words, the momentum 
constraint $C[0,N^a]$ may still be expected to be the Poisson bracket 
of ${\rm Im}\,Y$ and some real $W[N^a]$, or to be ${\rm Im}\,\{Y,W\}$ 
for some complex $W$. However, contrary to expectations, the Poisson 
bracket of $Y$ with $T[f]$ is not only the momentum constraint. It is 

$$\eqalign{
3\Bigl\{{\rm Re}\,Y,T[f]\Bigr\}&=-8\kappa\int_\Sigma f\tilde p^{ab}
  H_{ab}{\rm d}^3x=-{16\kappa^2\over3}\int_\Sigma{f\over\sqrt{\vert q
  \vert}}\varepsilon^{cda}\bigl(D_c\tilde p^b{}_d\bigr)\tilde p_{ab}\,
  {\rm d}^3x,  \cr
3\Bigl\{{\rm Im}\,Y,T[f]\Bigr\}&=8\kappa\int_\Sigma f\tilde p^{ab}\,
  {}_0E_{ab}{\rm d}^3x+{4\kappa^2\over3}\int_\Sigma{f\over\sqrt{\vert q
  \vert}}\tilde p\,\tilde{\cal C}\,{\rm d}^3x-2\kappa\, C[0,D^af]. \cr} 
\eqno(5.3)
$$
\ni
Therefore, {\it the constraints of the vacuum Einstein theory $C[N,
N^a]=0$ are equivalent to the conditions $\{{\rm Im}\,Y,V[N]\}=0$ 
and $3\{{\rm Im}\,Y,T[f]\}=8\kappa\int_\Sigma f\tilde p^{ab}\,{}_0
E_{ab}{\rm d}^3x$ on the Hamiltonian vector field of ${\rm Im}\,Y$, 
i.e. ${\rm Im}Y$ is constant along the volume flow, and varies in a 
specific way along the flow of $T[f]$}. It is not clear whether the 
momentum constraint function can also be written as the Poisson 
bracket of $Y$ (or of ${\rm Im}\,Y$) and some other function on 
$\Gamma_{ADM}$ (whenever the momentum constraint could also be 
interpreted as the condition of the invariance of $Y$ (or of ${\rm 
Im}\,Y$) along the flow of another Hamiltonian vector field), or not. 

\bigskip
\bigskip
\ni
{\lbf 6. The spinor Chern--Simons functional on the Ashtekar phase 
space}\par
\bigskip
\ni
By the triviality of ${\bf S}^A(\Sigma)$ the bundle of symmetric 
unprimed spinors is also globally trivializable: ${\bf S}^{(AB)}
(\Sigma)\approx\Sigma\times{\bf C}^3$, and if $\{\varepsilon^A_{\uA}\}$ 
is a global spin frame field in ${\bf S}^A(\Sigma)$, then $\varepsilon
^{AB}_{\bi}:=\sigma^{\uA\uB}_{\bi}\varepsilon^A_{\uA}\varepsilon^B
_{\uB}$, ${\bi}=1,2,3$, is a global frame field in ${\bf S}^{(AB)}
(\Sigma)$ and orthonormal with respect to the natural scalar product 
$\langle w^{AB},z^{AB}\rangle:=w^{AB}z^{CD}\varepsilon_{AC}\varepsilon
_{BD}$. For fixed bundle injection $\Theta^{AA'}_a$ the $SU(2)$ 
soldering form $\Theta^{AB}_a$ defines a (non-canonical) bundle 
isomorphism $\Theta:T\Sigma\otimes{\bf C}\rightarrow{\bf S}^{(AB)}
(\Sigma)$, and, in particular, the global frame field $\varepsilon^{AB}
_{\bi}$ can be written as $\varepsilon^{AB}_{\bi}=E^a_{\bi}\Theta^{AB}
_a$ for some globally defined complex basis $\{E^a_{\bi}\}$ in $T\Sigma
\otimes{\bf C}$. By the definitions, $\{E^a_{\bi}\}$ is orthonormal 
with respect to $q_{ab}$. However, the basis $\{E^a_{\bi}\}$ should not 
be confused with the orthonormal basis $\{e^a_{\bi}\}$ of $T\Sigma$ 
used in Section 2. The former is in fact a (complex) basis of ${\bf S}
^{(AB)}(\Sigma)$ in a disguise. Its densitized form, defined in a local 
coordinate system by $\tilde E^a_{\bi}:=\sqrt{\vert q\vert}E^a_{\bi}$, 
is therefore a triad of complex vectors of weight one. 
The connection 1-form and its curvature on ${\bf S}^{(AB)}(\Sigma)$ in 
this basis can be represented by $A^{\bi}_a:={1\over2}\varepsilon^{\bi}
{}_{\bj\bk}A^{\bj\bk}_a:={\rm i}\sqrt{2}\sigma^{\bi}_{\uA\uB}\Gamma
^{\uA\uB}_a$ and  $F^{\bi}{}_{ab}:={1\over2}\varepsilon^{\bi}{}_{\bj
\bk}F^{\bj\bk}{}_{ab}:={\rm i}\sqrt{2}\sigma^{\bi}_{\uA\uB}F^{\uA\uB}
{}_{ab}$, respectively. The latter can also be given by $F_{\bi\bj}:=
F_{{\bi}ab}E^a_{\bk}E^b_{\bl}\varepsilon^{\bk\bl}{}_{\bj}$, whose 
complete irreducible decomposition into its trace, anti-symmetric- and 
trace--free symmetric parts, expressed by the initial data of the 
second and third sections, is 

$$\eqalign{
F_{\bi\bj}=&-{1\over3}\,F^{\bk\bl}{}_{ab}\, E^a_{\bk}E^b_{\bl}\eta
  _{\bi\bj}-F^{\bl}{}_{ab}\, E^a_{\bl}E^b_{\bk}\varepsilon^{\bk}{}
  _{\bi\bj}+F_{\langle{\bi\bj}\rangle}=\cr
=&-{1\over3}\bigl(R+V\bigr)\eta_{\bi\bj}-{\rm i}D_b\bigl(\chi^b{}_a-
  \delta^b_a\chi\bigr)E^a_{\bk}\varepsilon^{\bk}{}_{\bi\bj}-2\bigl(
  {}_0E_{ab}+{\rm i}H_{ab}\bigr)\,E^a_{\bi}E^b_{\bj}.\cr}\eqno(6.1)
$$
\ni
Note that for given $\tilde E^a_{\bi}$ and $A^{\bi}_a$ the ADM 
canonical variables $q_{ab}$ and $\tilde p^{ab}$ are uniquely 
determined. 

The Ashtekar phase space $\Gamma_A$ is defined to be the set of the 
pairs $(A^{\bi}_a,\tilde E^a_{\bi})$ of the $so(3,{\bf C})$-valued 
connection forms and the complex triads of weight one, endowed with the 
symplectic structure whose restriction to the domain of the Ashtekar 
map ${\tt A}:\Gamma_A\rightarrow\Gamma_{ADM}:$ $(A^{\bi}_a,\tilde E^a
_{\bi})\mapsto(q_{ab},\tilde p^{ab})$ above is just the symplectic 
structure pulled back from $\Gamma_{ADM}$ along ${\tt A}$. This 
symplectic structure is ${\rm i}\kappa$-times the natural symplectic 
structure of $\Gamma_A$: If $A[\tilde\omega]:=\int_\Sigma A^{\bi}_a
\tilde\omega^a_{\bi}{\rm d}^3x$ and $E[\varphi]:=\int_\Sigma\tilde E
^a_{\bi}\varphi^{\bi}_a{\rm d}^3x$, the basic field variables smeared 
by arbitrary test fields with appropriate weights, then their Poisson 
bracket is $\{E[\varphi],A[\tilde\omega]\}={\rm i}\kappa\int_\Sigma
\varphi^{\bi}_a\tilde\omega^a_{\bi}{\rm d}^3x$. As a consequence of 
the non-injectivity of ${\tt A}$, i.e. the extra (internal) gauge 
freedom coming from the use of the triads $\tilde E^a_{\bi}$ instead 
of the metrics $q_{ab}$, a further constraint, the so-called Gauss
constraint $\tilde{\cal G}_{\bi}=0$, emerges in $\Gamma_A$. This system 
of constraints is 

$$\eqalignno{
\tilde{\cal S}&:=-{1\over2\kappa}{1\over\sqrt{\vert\det(\tilde E)
  \vert}}\,F^{\bi\bj}{}_{ab}\tilde E^a_{\bi}\,\tilde E^b_{\bj}=0,
  &(6.2.s)\cr
\tilde{\cal V}_b&:=-{1\over\kappa}\,F^{\bi}{}_{ab}\,\tilde E^a_{\bi}
  =0, &(6.2.v)\cr
\tilde{\cal G}_{\bi}&:=-{1\over\kappa}\,{\cal D}_a\,\tilde E^a_{\bi}
  =0; &(6.2.g)\cr}
$$
\ni
where, on the domain of ${\tt A}$ (i.e. on the `ADM-sector'), the 
constraints $\tilde{\cal S}$ and $-{\rm i}\tilde{\cal V}_a$ are 
precisely the pull backs to $\Gamma_A$ of $\tilde{\cal C}$ and $\tilde
{\cal C}_a$ along ${\tt A}$, respectively. The corresponding constraint 
functions on $\Gamma_A$ are $C[N,N^a,N^{\bi}]:=\int_\Sigma(\tilde{\cal 
S}N+(\tilde{\cal V}_a-\tilde{\cal G}_{\bi}\,A^{\bi}_a)N^a+\tilde{\cal 
G}_{\bi}N^{\bi}){\rm d}^3x$, where $N$, $N^a$ and $N^{\bi}$ are 
arbitrary real valued smearing fields on $\Sigma$. However, (6.2) on 
$\Gamma_A$ defines only the constraint system for the {\it complex} 
rather than the real, Lorentzian general relativity. To recover the 
latter, the so-called reality conditions, a further constraint, have 
to be imposed (see [9]). One of these conditions is ${\rm Im}\,(\tilde 
E^a_{\bi}\tilde E^b_{\bj}\eta^{\bi\bj})=0$, which, together with the 
implicit assumption $\det(\tilde E)\not=0$ that we had in (6.2.s), 
implies that $q^{ab}:=\vert\det(\tilde E)\vert^{-1}\tilde E^a_{\bi}
\tilde E^b_{\bj}\eta^{\bi\bj}$ is nondegenerate and negative definite, 
as it is on the ADM-sector. The other part of the reality conditions 
will not be used in what follows.

The spinor Chern--Simons functional $Y$ will now be a function 
of the configuration variable $A^{\bi}_a$ alone, and its functional 
derivative is ${\delta Y/\delta A^{\bi}_a}=F_{{\bi}bc}\epsilon^{bca}$, 
where $\epsilon^{abc}$ is the alternating Levi-Civita {\it symbol}. (In 
fact, $Y$ is just one-fourth the Chern--Simons functional $Y_{(2,0)}$ 
built from the connection $A^{\bi}_a$ on ${\bf S}^{(AB)}(\Sigma)$.) In 
the symplectic formalism its diffeomorphism- and gauge invariance are 
expressed by $\{C[0,N^a,0],Y\}=0$ and $\{C[0,0,N^{\bi}],Y\}=0$, 
respectively, which can also be verified directly (using the Bianchi 
identity for $F^{\bi}{}_{ab}$ in the latter case). Its Poisson bracket 
with $-C[N,0,0]$ coincides with $\dot Y$ given by (3.2), provided the 
constraints are satisfied. Thus in the generic case the Hamiltonian 
vector field of $Y$ is not tangent to the constraint surface.

On the ADM-sector the conformal rescaling of the previous section 
yields the transformation $\tilde E^a_{\bi}\mapsto\Omega^2\,\tilde E^a
_{\bi}$ and $A^{\bi}_a\mapsto\,A^{\bi}_a-\Omega^{-1}(\varepsilon^{\bi
\bj}{}_{\bk}E^b_{\bj}D_b\Omega+{\rm i}\delta^{\bi}_{\bk}\dot\Omega)
\vartheta^{\bk}_a$, and hence we define the conformal rescaling on the 
whole $\Gamma_A$ by the same formulae. In terms of the basic variables 
$E^a_{\bi}$ is defined by $\tilde E^a_{\bi}=:\sqrt{\vert\det(\tilde E)
\vert}E^a_{\bi}$, and $\vartheta^{\bi}_a$ is the dual of $E^a_{\bi}$. 
Thus the vector field on $\Gamma_A$ corresponding to the infinitesimal 
conformal rescaling by $(\delta\Omega,\delta\dot\Omega)$ is ${\cal K}=
(-\varepsilon^{\bi\bj}{}_{\bk}E^b_{\bj}(D_b\delta\Omega)\vartheta^{\bk}
_a-{\rm i}\delta\dot\Omega\vartheta^{\bi}_a,2\delta\Omega\,\tilde E^a
_{\bi})$. Then ${\cal K}(Y)=2{\rm i}\kappa C[\delta\dot\Omega,-{\rm i}
D^a\delta\Omega,{\rm i}A^{\bi}_aD^a\delta\Omega]$. ${\cal K}$ is a 
Hamiltonian vector field only if $\delta\Omega=0$, whenever the 
corresponding generator function is $\Phi=-{2\over\kappa}V[\delta\dot
\Omega]$, where $V[n]:=\int_\Sigma n\sqrt{\vert\det(\tilde E)\vert}{\rm 
d}^3x$. {\it Thus the scalar constraint $\tilde{\cal S}$ smeared by $n$ 
is just the Poisson bracket of $V[n]$, a functional of the momenta 
alone, and the spinor Chern--Simons functional $Y$, a functional of the 
configuration variable only; and $\tilde{\cal S}=0$ is a consequence of 
the requirement of the invariance of the spinor (or Ashtekar-) 
Chern--Simons functional with respect to infinitesimal spacetime 
conformal rescalings.} Since in the Ashtekar formulation $A^{\bi}_a$ 
does not have a metric content, {\it it is only $V[n]$ (i.e. the 
physical 3-volume or Misner's time if the reality conditions are 
satisfied) through which the spatial metric enters the dynamics}.

$T[f]$, the smeared version of York's local time, can be considered 
as a function on $\Gamma_A$, too, and its Poisson bracket with $Y$ is 
$\{Y,T[f]\}=-{1\over3}{\rm i}\int_\Sigma fF_{{\bi}[ab}(\Gamma^{\bi}
_{c]}-A^{\bi}_{c]})-{2\over3}{\rm i}\kappa\int_\Sigma\tilde{\cal V}_a
E^a_{\bi}\eta^{\bi\bj}E^b_{\bj}D_bf\,{\rm d}^3x$, which is precisely 
(5.3), where $\Gamma^{\bi}_a:={1\over2}\varepsilon^{\bi}{}_{\bj}{}
^{\bk}\vartheta^{\bj}_bD_aE^b_{\bk}$, the connection 1-form of the 
Levi-Civita connection determined by $E^a_{\bi}$ as an orthonormal 
basis. Interestingly enough, a similar formula for the vector 
constraint can be obtained by means of $M[f]:=\int_\Sigma f\Gamma
^{\bi}_a\tilde E^a_{\bi}{\rm d}^3x$, which is a (gauge dependent) 
function of the{\it momentum variable only}: $\{Y,M[f]\}=-{\rm i}
\kappa\int_\Sigma fF_{{\bi}[ab}\Gamma^{\bi}_{c]}-{\rm i}\kappa^2
\int_\Sigma\tilde{\cal V}_aE^a_{\bi}\eta^{\bi\bj}E^b_{\bj}D_bf\,{\rm 
d}^3x$. However, it is not clear whether the diffeomorphism and Gauss 
law constraints themselves can be recovered as the pure Poisson 
bracket of (the diffeomorphism and gauge invariant) $Y$ and some 
functionally differentiable functions $D,G:\Gamma_A\rightarrow{\bf 
C}$, too, or $Y$ generates only the constraint for the time 
evolution, but not for the kinematical symmetries. 

\bigskip
\bigskip

\ni
{\lbf Acknowledgments}\par
\bigskip
\ni
The author is grateful to Lars Andersson, Julian Barbour, Robert Beig, 
J\"org Frauendiener, Niall \'O Murchadha and Paul Tod for useful 
conversations, valuable remarks and stimulating questions. This work 
was partially supported by the Hungarian Scientific Research Fund 
grant OTKA T030374. 
\bigskip

\ni
{\lbf References}\par
\bigskip
\item{[1]} A. Ashtekar, R. Geroch, Quantum theory of gravitation, 
           Rep. Prog. Phys. {\bf 37} 1211 (1974)
\item{[2]} A. Ashtekar, {\it New Perspectives in Canonical Gravity}, 
           Bibliopolis, Naples, 1988
\item{[3]} J. Ehlers, H. Friedrich (Eds.), {\it Canonical Gravity: From 
           Classical to Quantum}, Springer-Verlag, Berlin 1994

\item{[4]} R. Haag, {\it Local Quantum Physics, Fields, Particles, 
           Algebras}, Springer Verlag, Berlin 1992
\item{[5]} R.M. Wald, {\it Quantum field theory in curved spacetime and 
           black hole thermodynamics}, University of Chicago Press, 
           Chicago, 1994
\item{[6]} C.J. Isham, Conceptual and geometrical problems in quantum 
           gravity, Imperial preprint, TP/90-91/14 
\item{   } C.J. Isham, Canonical quantum gravity and the problem of time, 
           Imperial preprint TP/91-92/25, gr-qc/9210011 
\item{   } C.J. Isham, Prima facie questions in quantum gravity, in [3]

\item{[7]} W.G. Unruh, R.M. Wald, Time and the interpretation of 
           canonical quantum gravity, Phys. Rev. D. {\bf 40} 2598 (1989)
\item{[8]} C. Rovelli, Quantum mechanics without time: A model, Phys. 
           Rev. D. {\bf 42} 2638 (1990)
\item{   } C. Rovelli, Time in quantum gravity: A hypothesis, Phys. Rev. 
           D. {\bf 43} 442 (1991)
\item{[9]} A. Ashtekar, {\it Lectures on Non-perturbative Canonical 
           Gravity}, World Scientific, Singapore 1991

\item{[10]} V.I. Arnold, {\it The mathematical methods of classical 
            mechanics}, M\H uszaki K\"onyvkiad\'o, Budapest 1985 (in 
            Hungarian)

\item{[11]} C.W. Misner, Mixmaster universe, Phys. Rev. Lett. {\bf 22} 
            1071 (1969)
\item{   }  C.W. Misner, Quantum cosmology. I, Phys. Rev. D. {\bf 186} 
            1319 (1969)
\item{[12]} J.W. York, Role of conformal three-geometry in the dynamics 
            of gravitation, Phys. Rev. Lett. {\bf 28} 1082 (1972)
\item{[13]} L. Smolin, C. Soo, The Chern--Simons invariant as the natural 
            time variable for classical and quantum cosmology, Nucl. Phys. 
            {\bf B449} 289 (1995)
\item{[14]} L.B. Szabados, On certain global conformal invariants and 
            3-surface twistors of initial data sets, Class. Quantum Grav. 
            {\bf 17} 793 (2000)
\item{[15]} J.B. Barbour, Timelessness of quantum gravity: I. The evidence 
            from the classical theory, Class. Quantum Grav. {\bf 11} 2853 
	     (1994)
\item{    } J.B. Barbour, Timelessness of quantum gravity: II. The 
            appearance of dynamics in static configurations, Class. 
	     Quantum Grav. {\bf 11} 2875 (1994)

\item{[16]} S.S. Chern, J. Simons, Characteristic forms and geometrical 
            invariants, Ann. Math. {\bf 99} 48 (1974)

\item{[17]} J.W. York, Gravitational degrees of freedom and the initial 
            value problem, Phys. Rev. Lett. {\bf 26} 1656 (1971)
\item{    } Y. Choquet-Bruhat, J.W. York, {\it The Cauchy problem}, in 
            General Relativity and Gravitation, Ed. A. Held, Plenum 
	     Press, New York (1980)
\item{    } J. Isenberg, Constant mean curvature solutions of the 
            Einstein constraint equations on closed manifolds, Class. 
	     Quantum Grav. {\bf 12} 2249 (1995)
\item{    } J. Isenberg, A set of nononstant mean curvature solutions 
            of the Einstein constraint equations on closed manifolds, 
	     Class. Quantum Grav. {\bf 13} 1819 (1996)

\item{[18]} R. Beig, L.B. Szabados, On a global conformal invariant of 
            initial data sets, Class. Quantum Grav. {\bf 14} 3091 (1997)

\item{[19]} R. Penrose, W. Rindler, {\it Spinors and spacetime}, vol 1, 
            and vol 2, Cambridge University Press, Cambridge, 1982 and 
            1986
\item{   }  S.A. Hugget, K.P. Tod, {\it An introduction to twistor theory}, 
            Cambridge Univ. Press, Cambridge 1985
\item{[20]} S. Kobayashi, K. Nomizu, {\it Foundation of differential 
            geometry}, vol 1 and vol 2, Interscience, New York 1964 and 
            1968
\item{    } F.W Warner, {\it Foundations of differentiable manifolds and 
            Lie groups}, Springer-Verlag, New York 1983

\item{[21]} P.E. Parker, On some theorems of Geroch and Stiefel, 
            J. Math. Phys. {\bf 25} 597 (1984)

\item{[22]} R. Beig, N. \'O Murchadha, The Poincare group as the symmetry 
            group of canonical general relativity, Ann. Phys. {\bf 174} 
	     463 (1987)
\item{[23]} D.R. Brill, S. Deser, Instability of curved spaces in 
            general relativity, Commun. Math. Phys. {\bf 32} 291 (1973)

\item{[24]} P.R. Chernoff, J.E. Marsden, {\it Properties of infinite 
            dimensional Hamiltonian systems}, Lecture Notes in Mathematics 
	     No. 425, Springer-Verlag, Berlin 1974
\item{    } J.E. Marsden, {\it Lectures on geometric methods in 
            mathematical physics}, SIAM, Philadelphia, Pennsylvania 1981

\item{[25]} A.E. Fischer, V. Moncrief, The Einstein-flow, the 
            $\sigma$-constant and the geometrization of 3-manifolds, 
            Class. Quantum Grav. {\bf 16} L79 (1999)

\item{[26]} R.F. Baierlein, D.H. Sharp, J.A. Wheeler, Three-dimensional 
            geometry as carrier of information about time, Phys. Rev. 
	     {\bf 126} 1864 (1962)

\end
\item{[26]} J.D. Brown, J.W. York, Jacobi's action and the recovery of 
	    time in general relativity, Phys. Rev. D {\bf 40} 3312 
            (1989)
\item{[27]} L. Bombelli, W.E. Couh, R.T. Torrence, Time as a spacetime 
            four-volume and Ashtekar variables, Phys. Rev. D {\bf 44} 
            2589 (1991) 
\item{[28]} J.B. Barbour, N. \'O Murchadha, Classical and quantum 
            gravity on conformal superspace, gr-qc/9911071
\item{[29]} J.B. Barbour, B.Z. Foster, N. \'O Murchadha, Relativity 
            without relativity, gr-qc/0012089

$\{T[f],A[n_e,n]\}={2\over3}A[n_e,fn]$, and $T[{3\over2}]$ acts on $A[n
_e,n]$ as an identity.  $A[n_e,n]$ has no critical points on $\Gamma
_{ADM}$. Their time evolution of is: 

$$\eqalignno{
\dot V[n]&:=\Bigl\{H[N,N^a],V[n]\Bigr\}=V[\L_{\bf N}n]-{3\kappa\over2}
  T[Nn], &(4.3.a)\cr
\dot A[n_e,n]&:=\Bigl\{H[N,N^a],A[n_e,n]\Bigr\}=A\bigl[n_e,n\bigl({2
  \kappa N\over\sqrt{\vert q\vert}}{\tilde p^{ab}n_an_b\over\vert n_c
  n_dq^{cd}\vert}-(q^{ab}+{n^an^b\over\vert n_cn_dq^{cd}\vert})D_aN_b
  \bigr)\bigr], &(4.3.b)\cr
\dot T[f]&:=\Bigl\{H[N,N^a],T[f]\Bigr\}=C_h[fN]+T[\L_{\bf N}f]+{2\over3
  \kappa}V[f(D_eD^eN+RN)],  &(4.3.c)\cr}
$$
\ni
For the area of a 2-surface ${\cal S}$ (4.3.b) gives $-\oint_{\cal S}(\nu 
k_{ab}+N\chi_{ab})r^{ab}{\rm d}{\cal S}$, where $r_{ab}$ is the induced 
2-metric on ${\cal S}$, $k_{ab}$ is the extrinsic curvature and $\nu$ 
is the outward normal component of $N^a$ with respect to ${\cal S}$. \par